\documentclass{article}

\usepackage{arxiv}

\usepackage[authoryear,longnamesfirst]{natbib}
\usepackage[utf8]{inputenc} % allow utf-8 input
\usepackage[T1]{fontenc}    % use 8-bit T1 fonts
\usepackage{hyperref}       % hyperlinks
\usepackage{url}            % simple URL typesetting
\usepackage{booktabs}       % professional-quality tables
\usepackage{amsfonts}       % blackboard math symbols
\usepackage{nicefrac}       % compact symbols for 1/2, etc.
\usepackage{microtype}      % microtypography
\usepackage{lipsum}
\usepackage{graphicx}
\usepackage{booktabs}
\usepackage{amsmath}
\usepackage{algorithm,algpseudocode}
\usepackage{algorithmicx}
\usepackage{makecell}
\usepackage{mathtools}
\usepackage{enumerate}
\usepackage{float}

\newcommand\blfootnote[1]{%
  \begingroup
  \renewcommand\thefootnote{}\footnote{#1}%
  \addtocounter{footnote}{-1}%
  \endgroup
}

\graphicspath{ {./images/} }

\DeclareUnicodeCharacter{2212}{-}

\title{MCD: A Modified Community Diversity Approach for Detecting Influential Nodes in Social Networks}

\author{
 Aaryan Gupta\footnotemark[1] \\
  Department of Applied Mathematics\\
  Delhi Technological University\\
  New Delhi, India\\
  \texttt{aryan227227@gmail.com} \\  
  \And
 Inder Khatri\footnotemark[1] \\
  Biometric Research Laboratory\\
  Delhi Technological University\\
  New Delhi, India\\
  \texttt{inderkhatri999@gmail.com} \\
  \And
 Arjun Choudhry \\
  Biometric Research Laboratory\\
  Delhi Technological University\\
  New Delhi, India\\
  \texttt{choudhry.arjun@gmail.com} \\
  \And
 Sanjay Kumar \\
  Department of Computer Science and Engineering\\
  Delhi Technological University\\
  New Delhi, India\\
  \texttt{sanjay.kumar@dtu.ac.in} \\
}

\begin{document}
\maketitle

\begin{abstract}
Over the last couple of decades, Social Networks have connected people on the web from across the globe and have become a crucial part of our daily life. These networks have also rapidly grown as platforms for propagating products, ideas, and opinions to target a wider audience. This calls for the need to find influential nodes in a network for a variety of reasons, including the curb of misinformation being spread across the networks, advertising products efficiently, finding prominent protein structures in biological networks, etc. In this paper, we propose Modified Community Diversity (MCD), a novel method for finding influential nodes in a network by exploiting community detection and a modified community diversity approach. We extend the concept of community diversity to a two-hop scenario. This helps us evaluate a node's possible influence over a network more accurately and also avoids the selection of seed nodes with an overlapping scope of influence. Experimental results verify that MCD outperforms various other state-of-the-art approaches on eight datasets cumulatively across three performance metrics.\blfootnote{*Equal Contribution}
\end{abstract}


\section{Introduction}\label{sec1}

Nowadays, Social networks have become an integral part of our daily lives and serve as an excellent medium for people to interact and propagate their views and ideas. Due to the large number of users available on various social networks across the globe, they act as expedient channels for advertisements and promotion of the products \citep{OSN1, OSN2, OSN3}. Over the years, advertisements over social networks, also known as “Viral Marketing”  have seen a significant boost due to their cost-effective nature and ability to propagate faster and wider. This approach, also known as “word-of-mouth advertising,” exploits people’s trust in a known or influential person’s suggestions over other traditional means of advertisement \citep{viralmarketingFerguson, VM}. Further, social networks have been used for various other uses, including building collaborations, friendships, and other social or professional relations. Hence, the problem of finding nodes with wide influence over a network is a prominent area of research in network analysis. This is further required for various tasks like curbing rumors spread in social networks and identifying prominent proteins in protein graphs and other graph-based networks.

The problem of finding influential nodes in a network can be delineated as the task of identifying a  set of $k$ influential nodes in a network whose influence over the whole network is maximum across any set of $k$ nodes, where $k$ is a small positive integral constant \citep{kempeIM,apin1}. Each individual entity acts as a node in the network, whereas the relationship between two nodes (friendship, follower-followee, co-authorship between users, links between proteins, etc.) helps form edges in the network to connect the nodes. For example, in social networks, the task entails maximizing the spread of an opinion by communicating it across the maximum number of nodes possible. Formally, the problem of finding the  influential nodes in a network, modelled as a graph $G(V, E)$  where $V$ is the set of all nodes in the network, and $E$ is the set of edges connecting the nodes, can be defined as shortlisting a small number of nodes ($k$)  such that the spread of influence can be maximized in the network. Mathematically, we can represent it as follows.
\begin{equation}
    S = argmax  \ \sigma(S) \, \ where \ S \in V , \ \|S\| = k, \|V\| \gg k
\end{equation}
Here, argmax  $sigma(S)$ denotes the overall influence spread generated by a set $S$ of $k$ number of nodes, and $k$ is a small and positive constant with respect to the total number of nodes in the network.
In general, selecting influential nodes in a network consists of two main components: choosing a set of $k$ spreaders or seed nodes and then computing the spread of influence across the network originating from the seed nodes utilizing a suitable information propagation model. While selecting the set of seed nodes, we first ensure that the number of seed nodes is significantly lesser than the number of nodes in the network to use the resources available optimally. We also need to consider the influence or propagating capability of each node in the network based on its locality and links within the network. Owing to the presence of a community structuring system in social networks, involving community-dependent features can be used to evaluate better the network structure, as well as aid in accelerating the spread of the propaganda.

In this work, we propose the concept of Modified Community Diversity (MCD) for finding influential nodes in a network. Our approach extends the idea of the previously proposed  Community-based Spreaders Ranking algorithm (CSR) \citep{CSR} for two-hop neighborhood information. The proposed work intends to enable better utilization of available seed nodes for information spread. We further propose the use of Extended Community Diversity for gathering information about the neighborhood based upon a two-hop mechanism. This information is further used to avoid the overlapping influence of influential nodes in a network, thus improving the efficacy of finding influential nodes in a network and providing an optimized spread of information or influence. For further experimental analysis, we compare the performance of our proposed approach with other state-of-the-art approaches for influential node detection on eight datasets, seven of which are real-world datasets. We use the Independent Cascade (IC) model \citep{IC} to simulate the influence propagation on all datasets.  Our experimental evaluations over three performance metrics validate the superiority of our proposed approach over the other approaches across different domains and act as evidence of its efficacy and widespread applicability. The main contributions of our paper can be summarised as follows.
\begin{itemize}
    \item We introduce Modified Community Diversity (MCD) method using community structure and extended community diversity to find influential nodes in social networks.
    
    \item The proposed work utilizes a two-hop concept which enables us to gather information about the neighborhood of a node better to judge its influence appropriately. This helps in avoiding the selection of influential nodes with overlapping influence.
    
    \item We verify the results of the proposed approach on eight network datasets against seven other state-of-the-art methods. The obtained results reveal that the proposed method outperforms different approaches cumulatively across three metrics. Statistical tests verify the superiority of MCD over competing approaches. 
\end{itemize}

The breakdown of the remaining paper is as follows. Section 2 details the related works and the existing research gaps. Section 3 elaborates on our proposed methodology and algorithms. Section 4 consists of information about the various experimental details used in our paper, including the datasets used, the Information Propagation models, the performance metrics, and the implementation environment. Section 5 contains our experimental results, their analysis, and broad outcomes. Section 6 delineates the statistical tests we have done to verify the efficacy of MCD for influence maximization. Section 7 concludes our paper. 

\section{Related Works}\label{sec2}

Due to the advent of the internet and the rise in the global usage of social networks, the task of finding influential spreaders in a network as a research problem has seen significant contributions from researchers. Researchers have proposed various approaches to find the most influential nodes in a network, to maximize the spread of information in a network \citep{IMsurvey, IMsurvey2}. These approaches can be predominantly categorized as diffusion process-based approaches, descriptive approaches, topological measures-based approaches, statistical and stochastic approaches, data mining-based approaches, and machine learning-based \citep{DL_based} approaches. Experimental evaluations have found topological measures-based approaches to be the most effective in identifying suitable spreader nodes \citep{MDER}. Topological measures-based approaches utilize predefined centrality measures for ranking nodes based on their efficacy in spreading information across a network. These centrality measures aid in selecting suitable spreader nodes with high influence on the other nodes in the network based on their specific topological properties. Centrality measures-based approaches generally consist of two main steps: generating a score for each node based on a defined centrality measure and selecting the top \emph{k} nodes based on the score obtained for the measure.

One of the first approaches using centrality scores for finding influential nodes in a network was proposed by \citet{Freeman}. The approach used the Degree Centrality measure, a local centrality measure that considers the number of neighbors of a node to generate its score. Degree Centrality has linear time complexity and was effective and efficient for small networks. However, due to its reliance on the local immediate neighbors for a given node, the effectiveness of Degree Centrality drops for nodes with few but influential neighbors. To counter the issues faced with local centrality measures, especially Degree Centrality, researchers proposed semi-local centrality measures like LC \citep{LC} and LSC \citep{LSC}, which go beyond the immediate scope of a node, but up to a limited range. While LC ranks the nodes using one-hop and two-hop nearest neighbors, LSC also considers the clustering coefficient between immediate neighbors. 

These local centrality-based approaches do not consider a node’s global location and only use the information corresponding to a certain locality, thus reducing their efficacy in real-life large networks. To counter this, researchers have also used global centrality measures like Betweenness Centrality, Closeness Centrality, and K-shell Decomposition, among others, to Influence Maximization. \citet{Betweenness} defined Betweenness for a specific node as the ratio of the shortest path which passes through the node and the total number of possible shortest paths in the network. \citet{Closeness} proposed Closeness for a node as the inverse of the sum of the shortest path to all other nodes, thus determining the average closeness of a node to the rest of the graph. Betweenness and Closeness suffer from high time complexity, as they require the calculation of the shortest path between every node pair of the network. 

Some other global centrality measures successfully used for Influence Maximization are Eigenvector Centrality \citep{Eigenvector} and Pagerank Centrality measure \citep{Pagerank}. They aid in ranking nodes based on their importance among the neighborhood nodes. According to these measures, the presence of important nodes around the central node makes it a viable spreader node. Another global centrality-based measure, GLR \citep{GLR}, divides the graph into communities and then selects each community's local cores and gateway nodes. The nodes are ranked as influential spreaders based on their shortest paths to the local cores and the gateway nodes. Researchers have also proposed greedy search-based approaches \citep{Greedy1, Greedy2} for ranking spreader nodes. These approaches use the sub-modularity and monotony of the Independent Cascade (IC) model along with Monte Carlo simulations to enhance the overall performance of the approach. However, greedy approaches are generally inefficient due to many simulations, and they require a lot of time to compute.

Some recent approaches also consider the bridge nodes, which connect two different groups of nodes in a network, as influential nodes. \citet{DIL} proposed the DIL centrality measure, where the importance of lines and the degree value are used to highlight a bridge node. DIL also considers the number of connected triangles when determining the relevance of a link, with the bridge node taking the primary role in controlling information propagation. \citet{DCL} introduced the DCL centrality measure, which finds the bridge nodes using a node’s degree, clustering coefficient, and the relationship between its one-hop neighbors. \citet{LID} introduced LID, an approach where prominent influencers in complex networks are extracted using quasi-local information, i.e., local structural properties.

Recently, researchers have proposed various community structure-based approaches \citep{communityJIIS, communityJIIS2} for spreader ranking, which were found to be more effective in identifying suitable spreader nodes in a network \citep{huang2019community, IM-ELPR, Community_based_IM}. Some recent community-based algorithms used for Influence Maximization are IM-ELPR \citep{IM-ELPR}, CFIN \citep{CFIN}, and CAOM \citep{CAOM}. CAOM divides the whole network into communities and selects only the most significant ones. Certain candidate nodes are then selected from each community using the one-hop measure. Hereafter, seed nodes are selected from candidate nodes by eliminating those with influence overlap. CFIN follows a similar approach, but the candidate nodes are selected using degree centrality, and the seed nodes are selected using the clustering coefficient measure. IM-ELPR uses node seeding, community detection, and label propagation for influence maximization. The seed nodes are first identified using an extended h-index, and then the communities are detected using label propagation. The smaller, related communities are further merged into larger communities, and then the top $k$ influential nodes are identified.

The approaches mentioned above utilized local structures to identify the bridge nodes. However, unlike these approaches, the recently proposed Community-based Spreaders Ranking algorithm (CSR) \citep{CSR} for influence maximization selects global bridge nodes by considering the community bridge nodes as the most influential spreaders in the network. The CSR algorithm utilized three measures to identify bridge nodes, i.e., community diversity, community modularity, and community density. However, CSR didn't consider the overlapping influence of the influential nodes in a network. Our proposed approach (MCD) considers this factor and thus enables better utilization of available seed nodes. Further, CSR considers only one-hop neighbors to detect influential nodes. In contrast, MCD uses a two-hop mechanism, which provides a better overview of the node's neighborhood provided to the algorithm. This leads to markedly superior results by our approach as compared to CSR.

\begin{figure*}[htp]
\centering
%\begin{center}
\centerline{\includegraphics[scale=0.90]{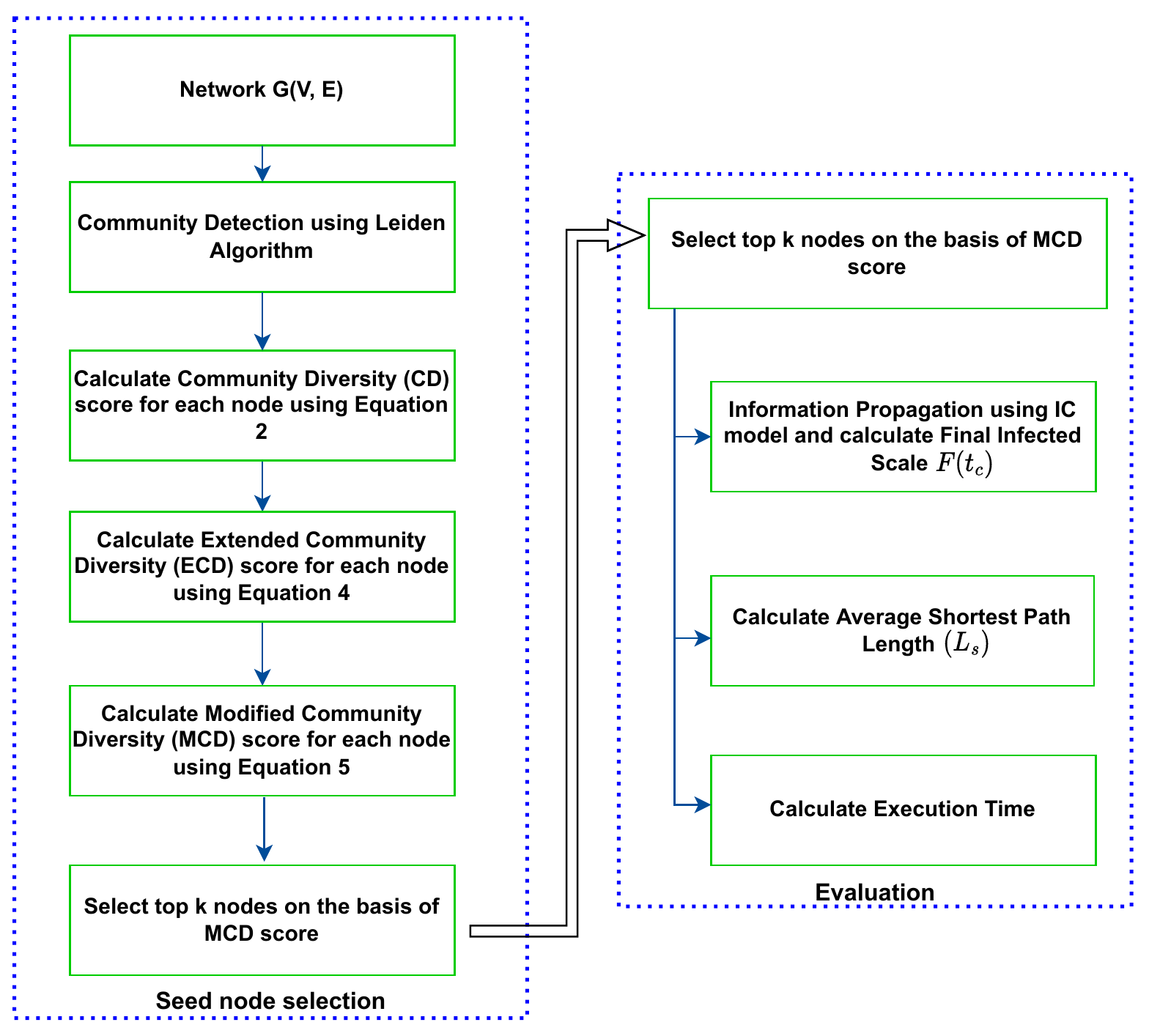}}
%\end{center}
%\scriptsize
%%%%%%%%%%%%%%%%
%\vspace{-0.1 in}
%%%%%%%%%%%%%%%%
\caption{Flowchart representation of the steps involved for the calculation of MCD for each node.} 
%makes training less sensitive to the frequency spectrum vulnerable to adversarial perturbations. We use low frequency images at both pre training and fine tuning stages and further do evaluation on low frequency samples.}
\label{fig:flowchart}
\end{figure*}

\section{Proposed Methodology}\label{sec3}

In this section, we present our proposed methodology for  finding influential nodes in social networks using community detection and a modified community diversity approach. First, we uncover the community structure in the input network  using Leiden's algorithm. We consider the bridge nodes in the network to play a vital role in spreading the information, and thus, we identify the bridge nodes in the network using the Community Diversity (CD) score. To incorporate the two-hop mechanism for gathering more information regarding a node's surroundings, we calculate the Extended Community Diversity (ECD) score. Then we calculate the Modified Community Diversity score using ECD and CD to ensure the proper distribution of the information. We rank the nodes based on the MCD score and select the top $k$ nodes, where $k$ signifies the constraint on the number of nodes that act as influencers. Finally, we evaluate the influence of the selected $k$ seed nodes as influential nodes over the entire network by using various evaluation parameters.
The overall proposed work can be understood using Fig. \ref{fig:flowchart}, which depicts the process flow for calculating the MCD score for each node.  The various phases of the proposed work are discussed below. 

\subsection{Community Detection}\label{community}

A community can be illustrated as a subset of a graph with nodes representing each member of the community \citep{yang2021spiderweb, community, communityJIISMetrics}. The nodes have similar characteristics and are connected to some of the other nodes by edges. These nodes can represent any entity in the world, and their meaning is dependent on the scope of the community and, to a major extent, the scope of the graph. Nodes present specifically in a community are expected to have meaningful and substantial connections with each other in comparison to the rest of the graph, thus ensuring a better spread of information within the community as compared to outside the community. In our proposed algorithm, we use Leiden’s algorithm \citep{Leiden} to identify the communities in a network. Leiden's algorithm is based on greedy optimization techniques, which results in efficient processing and quicker convergence. Leiden's algorithm consists of three steps:
\begin{itemize}
    \item the local moving of the nodes.
    \item refining the partition between communities.
    \item accumulation of the nodes in the network on the basis of the refined partition, with the non-refined partition creating an initial partition for the accumulated network.
\end{itemize}

\subsection{Community Diversity Score}
It is often observed in real-world networks that influential nodes are usually linked to multiple communities in the network. Thus, they have a significant influence over the state of the network, as they can simultaneously disseminate their influence to multiple communities. We consider these high-degree nodes, which are connected to multiple communities, as influential nodes by applying the concept of Community Diversity \citep{CSR}, as they act as bridge nodes between communities. The community diversity score of a node can be formally defined as the measure of the probability of finding a certain community in the neighborhood of a node. Community diversity gives a high score to a node that is connected to multiple communities, as well as considers the number of communities to which the node transmits information. The concept of community diversity revolves around the concept of Shannon's entropy \citep{Shannon} over the term of the probability of finding the $v^{th}$ node in the $c^{th}$ community in the neighborhood of a certain node.
The Community Diversity score for a node is computed as shown in Equation \ref{eq:cdv}, where $CDv_{i}$ is the community diversity value of the $i^{th}$ node, and $CDv_{ic}$ is the contribution of $c^{th}$ community in Community Diversity of $i^{th}$ node \citep{CSR}. The neighborhood contribution of each community is shown in Equation \ref{eq:p}, where $\eta_{in}$ represents the number of nodes in the $c^{th}$ community in the neighborhood of node $i$, and $\eta_{i}$ represents the number of nodes in the neighborhood of node $i$. 

\begin{equation}\label{eq:cdv}
CD_{ i } =-\sum_{c = 1}^{comm} CDv_{ ic } = - \sum_{c = 1}^{comm} P_{ ic } * log(P_{ ic })
\end{equation}

\begin{equation}\label{eq:p}
    P_{ic} = \frac{\eta_{ic}}{\eta_{i}}
\end{equation} 

\begin{algorithm}[h!]
	\caption{: Community Diversity Score \citep{CSR}}
	\begin{algorithmic}[1]
		\renewcommand{\algorithmicrequire}{\textbf{Input:}}
		\renewcommand{\algorithmicensure}{\textbf{Output:}}
		\Require  $G = (V,E)$, where $n=\|V\| , e = \|E\|$
		\Ensure  $Community\_Diversity$\\
		$Initialization$
		%\STATE $Community\_Diversity \leftarrow \{\}$ \\
		%\FOR {$v$ $\in$ $\{$1,2....n$\}$}
		\State $Communities\ \{ $c1, \ c2....cm$ \}:\ partitioned\ using\ Leiden's\ Algorithm $
		\State $value \leftarrow 0$
		\State $CD \leftarrow \{\}$
		\For {$node$ $\in$ $\{Nodes\ of\ Graph\ G\}$}
		\State $sum \leftarrow 0$
		\State $neigh\_len \leftarrow number\ of\ neighbours\ of\ node$
		\For {$c$ $\in$ $\{$Communities$\}$}
		\State $neigh\_com\_len \leftarrow number\ of\ neighbours\ of\ node\ belonging\ to\ c $
		\State $ratio \leftarrow \frac{neigh\_com\_len}{neigh_len}$
		\State $value \leftarrow - ratio*log(ratio)$
		\State $sum \leftarrow sum + value$
		\EndFor
		\State $CD(node) \leftarrow sum$
		\EndFor\\
		%\ENDFOR
		\Return $CD$
	\end{algorithmic}
	\label{algo:CD}
\end{algorithm}

Algorithm \ref{algo:CD} represents the procedure used to compute the Community Diversity (CD) score for a graph $G$. It can be summarized as follows:
\begin{itemize}
	\item We divide the graph into communities using Leiden's algorithm. 
	\item For each node, we calculate the number of neighbours. Further, for each community, the number of neighbours of a node that are part of the respective community is computed.
	\item We evaluate the ratio of the total number of neighbours of the node and the neighbour of the node in the respective community. This is further used to calculate the entropy score, i.e., the Community Diversity score.
\end{itemize}

If a node is connected to multiple communities and the connections are found to be uniformly distributed, then the community diversity score of that node is observed to be higher. However, if a node has connections to only a few communities or the distribution of connections is not uniform, then the community diversity score is observed to be low. Hence, the concept of community diversity can be utilized to find out the nodes capable of efficiently spreading information to multiple parts of the network.  

\subsection{Extended Community Diversity Score}
We formally propose the Extended Community Diversity score as the ratio of the community diversity score of a node to that of its neighborhood. The concept of Extended Community Diversity revolves around considering a node along with its neighborhood for the prediction of its influence in a locality. The neighbors of a node contribute to helping the algorithm understand what relative position a node holds with respect to its neighborhood, which ultimately tells how much a node will benefit from being made the seed node. Extended Community Diversity gains information about the neighborhood of a node by taking two hops, i.e., one hop from the node to its neighbors, and the second hop from its neighbors to farther nodes. This provides more information for the proposed algorithm to process and deduces which node is most influential. The provided information is further used to avoid having many influential nodes in the same region and hence avoid the wastage of useful resources.
Extended Community Diversity score for a node is computed as shown in Equation \ref{eq:ECD} where $ECD_{v}$ is the Extended Community diversity value of $v^{th}$ node, $neigh$ is the neighbours of node $v$ and $CD_{v}$ is the community diversity score for node $v$ as computed in Equation \ref{eq:cdv}.

\begin{equation}\label{eq:ECD}
ECD_{v} = CD_{v} + \sum_{n = 1}^{neigh} CD_{n}
\end{equation}
% \begin{equation}\label{eq:p}
%     P_{ic} = \frac{\eta_{ic}}{\eta_{i}}
% \end{equation} 

% Algorithm Density
\begin{algorithm}[h!]
 \caption{: Extended Community Diversity Score}
 \begin{algorithmic}[1]
 \renewcommand{\algorithmicrequire}{\textbf{Input:}}
 \renewcommand{\algorithmicensure}{\textbf{Output:}}
 \Require 
 $G = (V,E)$, where $n=\|V\| , \ e = \|E\|$
  \Ensure  $Extended\_Community\_Diversity$\\
  $Initialization$
  \State $ ECD \ \leftarrow \ \{\} $ 
  \State $ CD \ \leftarrow \ Community \ Diversity \ Score \ using \ algorithm \ 1 $
  \State $ value \ \leftarrow \ 0$
  \For { $v \in$ $ \{Nodes\ of\ Graph\ G\} $ }
    \State  $value \ \leftarrow CD(v)$
    \For {$nb$ $\in$ $\{Neighbours\ of\ node\ v\}$}
        \State $value \leftarrow   value$ + $CD(nb)$ 
    \EndFor
    \State $ ECD(v) \leftarrow   value$ 
  \EndFor\\
\Return $ECD$
\end{algorithmic}
\label{algo:ECD}
\end{algorithm}

Algorithm \ref{algo:ECD} represents the algorithmic procedure used to compute the Extended Community Diversity (ECD) score for graph G. The summary of the algorithm is as follows:
\begin{itemize}
% Community K-shell refers to the concept where each community is treated as a separate graph and K-Shell algorithm executes over each community separately.
\item We calculate the previously defined Community Diversity score (CD) in Algorithm \ref{algo:CD}.
\item Then, for each node, we calculate the summation of its Community Diversity score, and its neighbor's Community Diversity Score, which gives the Extended Community Diversity score.
\end{itemize}

\subsection{Modified Community Diversity Score}
We formally define the Modified Community Diversity (MCD) score as $MCD = -P*logP$, where $P$ is the ratio of the Community Diversity of a node to its Extended Community Diversity. Modified Community Diversity uses the limited available resources judiciously by ensuring that no two influential nodes are in the same locality, which, if not taken care of, would result in the wastage of valuable spreading power and overlapping of the region of influence of influential nodes in the same locality. The problem of several influential nodes existing in the same neighborhood is known as the Rich Club effect. Our proposed approach tackles this problem by scoring the nodes based on their strength relative to their neighboring nodes, thus ensuring that the influential nodes are well spread throughout the network. The ratio of the results of the previously calculated CD score and ECD score is used for analyzing the MCD score. The concept of MCD maximizes when the distribution of spreading power is uniform, which makes sense here since the algorithm not only wants the respective node to be influential but also the neighborhood to be sufficiently influential in carrying on the spread of information further. This enables the node to maximize the spread of information in that locality and beyond it. This results in the respective node being somewhat equivalent to the whole neighborhood in terms of spreading power, thus making it the most influential node in the locality. This mitigates the problem of the rich club effect significantly by considering the overlap in the influence and handling it in a balanced way to avoid wastage of useful spreading strength. This results in an optimized spread of information with assured maximum throughput. Modified Community diversity score for a node is computed as shown in Equation \ref{eq:pMCD}, where $P_{v}$ is as shown in Equation \ref{eq:pMCD}. In Equation \ref{eq:pMCD}, $CD_{v}$ is the community diversity score of $v^{th}$ node, as shown in Equation \ref{eq:cdv}, and $ECD_{v}$ is the Extended Community diversity value of $v^{th}$ node, as calculated in Equation \ref{eq:ECD}.

\begin{equation}\label{eq:MCD}
MCD_{v} = - P_{v} * log(P_{v})
\end{equation}

\begin{equation}\label{eq:pMCD}
    P_{v} = \frac{CD_{v}}{ECD_{v}}
\end{equation}

% EQUATION
% P = CD(n)/ECD(n)
% Val = -P*Log(P)

%ALGORITHM 2 MODULARITY
\begin{algorithm}[h!]
 \caption{: Modified Community Diversity Score}
 \begin{algorithmic}[1]
 \renewcommand{\algorithmicrequire}{\textbf{Input:}}
 \renewcommand{\algorithmicensure}{\textbf{Output:}}
 \Require $G = (V,E)$, where $n=\|V\| , e = \|E\|$ 
  \Ensure  $Modified\_Community\_Diversity$\\
  $Initialization$
  \State $ $CD $ \leftarrow \ $Community Diversity Score using algorithm 1$ $
  \State $ $ECD $ \leftarrow \ $Extended Community Diversity Score using algorithm 2$ $
  \State $MCD \leftarrow \{\}$ 
  \State $value \leftarrow 0$
  \State $p \leftarrow 0$
  \For {$v$ $\in$ $\{Nodes\ of\ Graph\ G\}$}
    \State $p \leftarrow \frac{CD(v)}{ECD(v)}$
    \State $value \leftarrow -p*log(p)$
    \State $MCD(v) \leftarrow value$ 
  \EndFor\\
 \Return $MCD$
 \end{algorithmic} 
 \label{algo:MCD}
\end{algorithm}

Algorithm \ref{algo:MCD} represents the procedure used to calculate the Modified Community Diversity (MCD) score for graph G. The summary of the algorithm is as follows:
\begin{itemize}
% Community K-shell refers to the concept where each community is treated as a separate graph and K-Shell algorithm executes over each community separately.
\item The community diversity score (CD) is calculated using Algorithm \ref{algo:CD}, and the extended community diversity score (ECD) is calculated using Algorithm \ref{algo:ECD}.
\item Then, for each node, we calculate the ratio of the Community Diversity score and Extended Community Diversity Score, which gives us the MCD score for the respective node. 
\end{itemize}

Figure \ref{fig:flowchart} illustrates the process flow for calculating the MCD score for each node. We first calculate the CD and ECD scores for each node. Using the generated values of CD and ECD scores, we compute the MCD scores for each node. We further rank the nodes on the basis of the MCD score and select the top $k$ nodes, where $k$ signifies the constraint on the number of nodes that act as influencers. We further evaluate the influence of the selected $k$ nodes over the entire network in Section \ref{sec5}.

\section{Datasets, Information Propagation Model, \& Performance Metrics}\label{sec4}

\subsection{Dataset and Baseline Models}
We evaluate our proposed approach on seven real-life datasets and one synthetic dataset of varying sizes and types. Table \ref{table-1} describes the different datasets used during our evaluation. Further, we do a comparative analysis of the proposed approach with seven other state-of-the-art approaches, namely CSR \citep{CSR}, GLR \citep{GLR}, DCL \citep{DCL}, LID \citep{LID}, DIL \citep{DIL}, H-Index (HI) \citep{H_Index} and PageRank (PR) \citep{Pagerank}, over three distinct evaluation criteria. For the optimal performance for the baseline approaches, we took the input parameters as reported in the original papers throughout all our experiments.

\begin{itemize}
    \item Community Spreaders Ranking (CSR) \citep{CSR}: CSR measures a node's spreading ability with respect to different communities on the basis of its connection to the communities.
    \item  Gateway Local Rank (GLR) \citep{GLR}: GLR extends over the closeness centrality measure and simplifies it by reducing the search set to the local and gateway nodes. To get the optimal performance, we take parameters $\alpha_1$ and $\alpha_2$ used for weighing the effect of core and gateway nodes to be 1 as reported by \citep{GLR}. 
    \item Degree and Clustering Coefficient and Location (DCL) \citep{DCL}: DCL measures the spreading ability of nodes on the basis of a node's degree, its neighbour, and the clustering coefficient.  
    \item Local Information Dimensionality (LID) \citep{LID}: LID measures the spreading ability of a node by considering the quasilocal structure of a node.
    \item Degree and Importance of Lines (DIL) \citep{DIL}: DIL ranks the nodes on the basis of the degree and the importance of lines.
    \item H-index (HI) \citep{H_Index}: H-index is a node centrality measure that considers the neighborhood measures, the quality, as well as the quantity of the neighboring nodes simultaneously.     
    \item PageRank (PR) \citep{Pagerank}: PageRank measures the significance of a given node in the network based on its outgoing degree and damping factor. Similar to the work \citet{Pagerank} and various other works, we take the value of $d$ to be $0.85$ in all experiments to get optimal performance.
\end{itemize}

\begin{table}[!ht]
    
    \centering
    \caption{The different datasets used for experimentation along with the various characteristics such as Number of Nodes, Edges, and Communities by Leiden Algorithm}
    \begin{tabular}{|l|l|l|l|l|l|}
    \hline
        S.No. & Dataset & Nodes (n) & Edges (m) & Communities \\ \hline
        1 & BA & 2,000 & 9,974 & 14\\ \hline
        2 & PGP & 10,638 & 24,301 & 104\\ \hline
        3 & Jazz Musicians & 198 & 2,742 & 4\\ \hline
        4 & Ca-hepth & 7,126 & 35,324 & 481\\ \hline
        5 & GR-QC & 5,242 & 14,496 & 394\\ \hline
        6 & Dolphins & 62 & 159 & 5\\ \hline
        7 & Erdos992 & 6,100 & 7,500 & 50\\ \hline
        8 & p2p-Gnutella25 & 22,687 & 54,705 & 59\\ \hline        
    \end{tabular}
    \\
\label{table-1}
\end{table}

A brief description of different datasets is given below:

\begin{itemize}
\item BA \citep{ba}: BA is a random graph generated using Barabási-Albert preferential attachment model. A network of \textit{n} nodes is generated by adding new nodes, each having \textit{m} edges that are preferentially coupled to existing nodes with a high degree. 
\item PGP \citep{PGP}:  Pretty Good Privacy (PGP) is an encrypted communication network. PGP algorithm is used for secure information interchange. Here, the users are nodes, and the edges of this huge network are the connectivity between users.
\item Jazz Musicians \citep{Jazz}: The Jazz musicians dataset is a real-world social network dataset, where each node represents a musician, and each edge between two nodes indicates that the two musicians have played together in a band.
\item Ca-hepth \citep{Ca-hepth}; It represents a collaboration network of authors who have collaborated on papers submitted to the High Energy Physics - Theory Category.
\item General Relativity and Quantum Cosmology (GR-QC) \citep{GRQC}: GR-QC dataset is a real-world social network dataset from arXiv and covers scientific contributions by authors in General Relativity and Quantum Cosmology. The authors are nodes, and if they have collaborated with another author on a paper, then there is an undirected edge between them.
\item Dolphins dataset \citep{dolphins}: The dolphins dataset is a social network of bottlenose dolphins, where a node represents a dolphin, while an edge represents frequent associations between dolphins.
\item Erdos992 \citep{Erdos992}: This dataset is about the collaborations of Paul Erdős, a Hungarian Mathematician, with his co-authors. The co-authors are nodes while collaborations between the nodes signify the edges.
\item p2p-Gnutella25 \citep{Gnutella}: This dataset is one of the snapshots of the Gnutella peer-to-peer file sharing network from August 2002. Nodes represent hosts in the Gnutella network, while edges represent connections between the Gnutella hosts.

\end{itemize}

\subsection{Information Propagation Model}

Information Propagation (IP) is described as the task of circulating information in complex networks \citep{IP1, IP2}. IP can be used to assess various spreader-ranking approaches. Generally, epidemic models are used to model information propagation in a complex network. In epidemic models, the population of users is represented as nodes, and the requisite information is regarded as an infectious disease, which propagates through the population upon the interaction between nodes.

While various epidemic models can be used for information propagation tasks, like SIR \citep{SIR}, IC \citep{IC} and LT \citep{LT}, we employ the Independent Cascade (IC) model to assess the performance of our proposed approach and evaluate it against other state-of-the-art approaches. The nodes in an IC model can be allocated one out of two states: infected node or susceptible node. The propagation of information in the network begins with setting the state of a few nodes as infected. These nodes further interact with their neighboring nodes and try to infect them. Further, the newly infected nodes further interact with their neighbors to further propagate the information to their neighbouring nodes until there are no new nodes in the network that are infected. A probability ($P$) is associated with these interactions between nodes, which elucidates the possibility of a susceptible node getting infected upon interaction with an infected node. The total influence of a seed node is calculated by taking the sum of all nodes which get infected by the end of information propagation.

\subsection{Performance Metrics}

We use three performance metrics for comparing our approach with other state-of-the-art approaches. They are as follows:

(i) Final Infected Scale: It is defined as the ratio of activated or infected nodes in the network to the total number of nodes in the network at the end of information propagation. The nodes only get activated when the activation probability for a node is favourable for activation. The fraction of the total nodes taken as the initial spreaders is called the spreader fraction, and it plays a significant role in the final number of nodes that get infected. Generally, a high spreader fraction leads to a high percentage of the total nodes getting infected.

(ii) Average Shortest Path Length (ASPL): ASPL can be defined as the average of the shortest paths between two influential nodes for all sets of influential nodes. It enables us to evaluate the efficacy of information propagation in a complex social network. The higher the value of ASPL, the more even the spread of the influential nodes in the network. A high value of ASPL further indicates that influential nodes cover a wider part of the network instead of being concentrated in a small region, leading to the poor spread of information in the network.

(iii) Execution Time: Execution Time: We evaluate our approach against other approaches based on the time taken to rank nodes based on their ability to propagate information in a network. The time consumption for all centralities was assessed in a Google Colaboratory environment. The code was run on an Intel Xeon CPU with one core, with 13GB of RAM available for use. 

\section{Experimental Results and Analysis}\label{sec5}

\begin{figure*}[h!]
        \centering
	% Use the relevant command to insert your figure file.
	% For example, with the graphicx package use
	\includegraphics[scale=0.48]{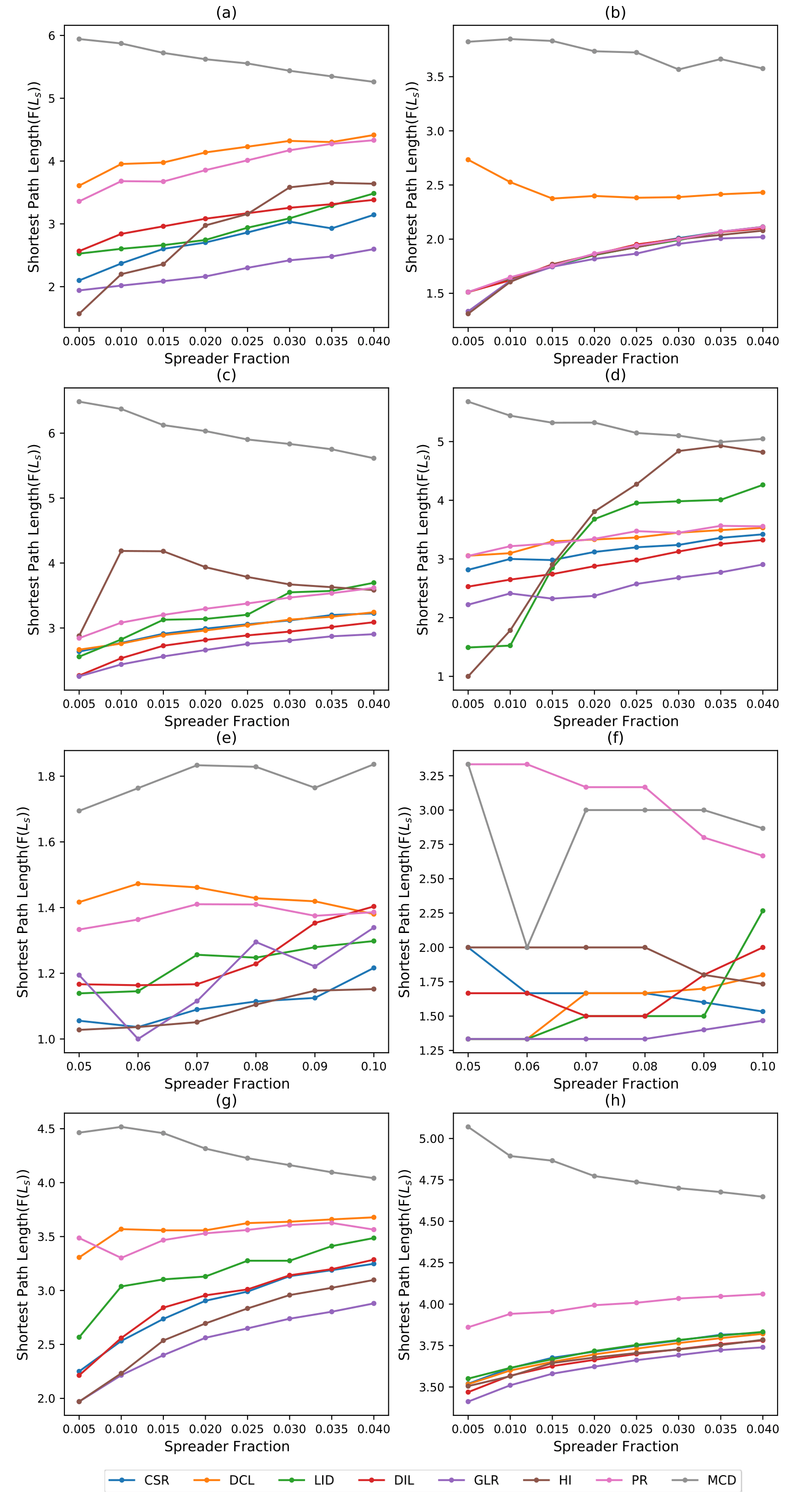}
	% figure caption is below the figure
	\caption{Average distance between spreaders with respect to different initial spreaders fraction values for (a) PGP dataset (b) BA dataset (c) Ca-hepth dataset (d) GR-QC dataset (e) Jazz Musicians dataset (f) Dolphins dataset (g) Erdos992 dataset (h) p2p-Gnutella25 dataset. The results are averaged for 100 independent simulations of the Information Cascade model with an activation probability (P) equal to 0.1.}
	\label{LS}       % Give a unique label      
\end{figure*}

\begin{figure*}[h!]
        \centering
	% Use the relevant command to insert your figure file.
	% For example, with the graphicx package use
	\includegraphics[scale=0.48]{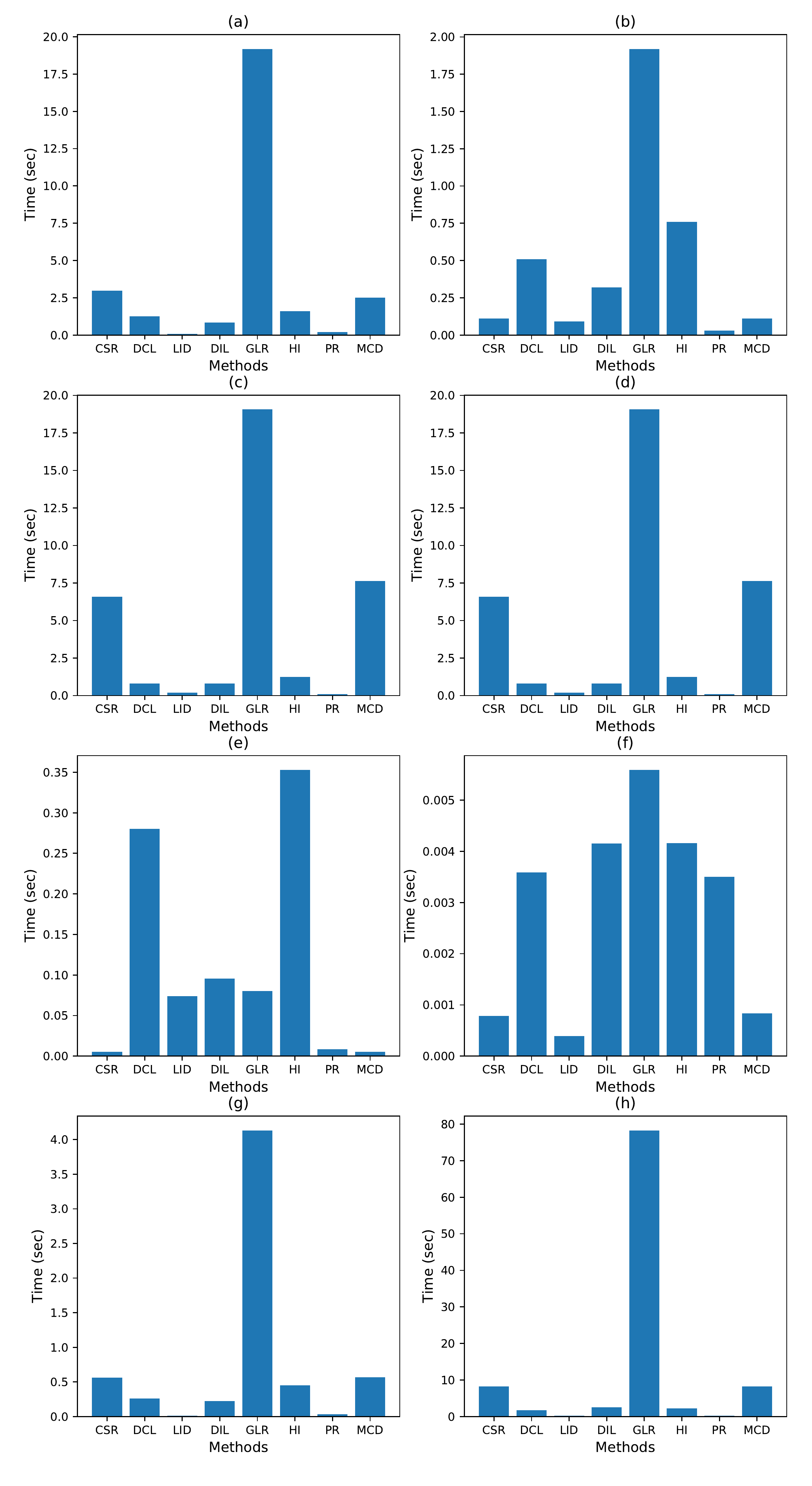}
	% figure caption is below the figure
	\caption{The execution time plots for different methods for ranking the spreaders for (a) PGP dataset (b) BA dataset (c) Ca-hepth dataset (d) GR-QC dataset (e) Jazz Musicians dataset (f) Dolphins dataset (g) Erdos992 dataset (h) p2p-Gnutella25 dataset.}
	\label{ET}       % Give a unique label      
\end{figure*}

This section presents the experimental results and analysis performed on the eight datasets using three performance metrics as mentioned in Section \ref{sec4}. We conducted a comparative analysis of the proposed approach with seven other state-of-the-art approaches for finding influential nodes, namely,  H-Index (HI) \citep{H_Index}, PageRank (PR) \citep{Pagerank}, Community-based Spreader Ranking (CSR) \citep{CSR}, Gateway Local Ranking (GLR) \citep{GLR}, Degree and Clustering Coefficient and Location (DCL) \citep{DCL}, Local Information Dimensionality (LID) \citep{LID}, and Degree and Importance of Lines (DIL) \citep{DIL}. We utilized the Information Cascade (IC) model to simulate information propagation. Our analysis of the results obtained over three metrics, illustrated in Figures \ref{FTC}, \ref{LS}, and \ref{ET}, is elaborated below. 

\subsection{Final Infected State}
\begin{figure*}[h!]
        \centering
	% Use the relevant command to insert your figure file.
	% For example, with the graphicx package use
	\includegraphics[scale=0.48]{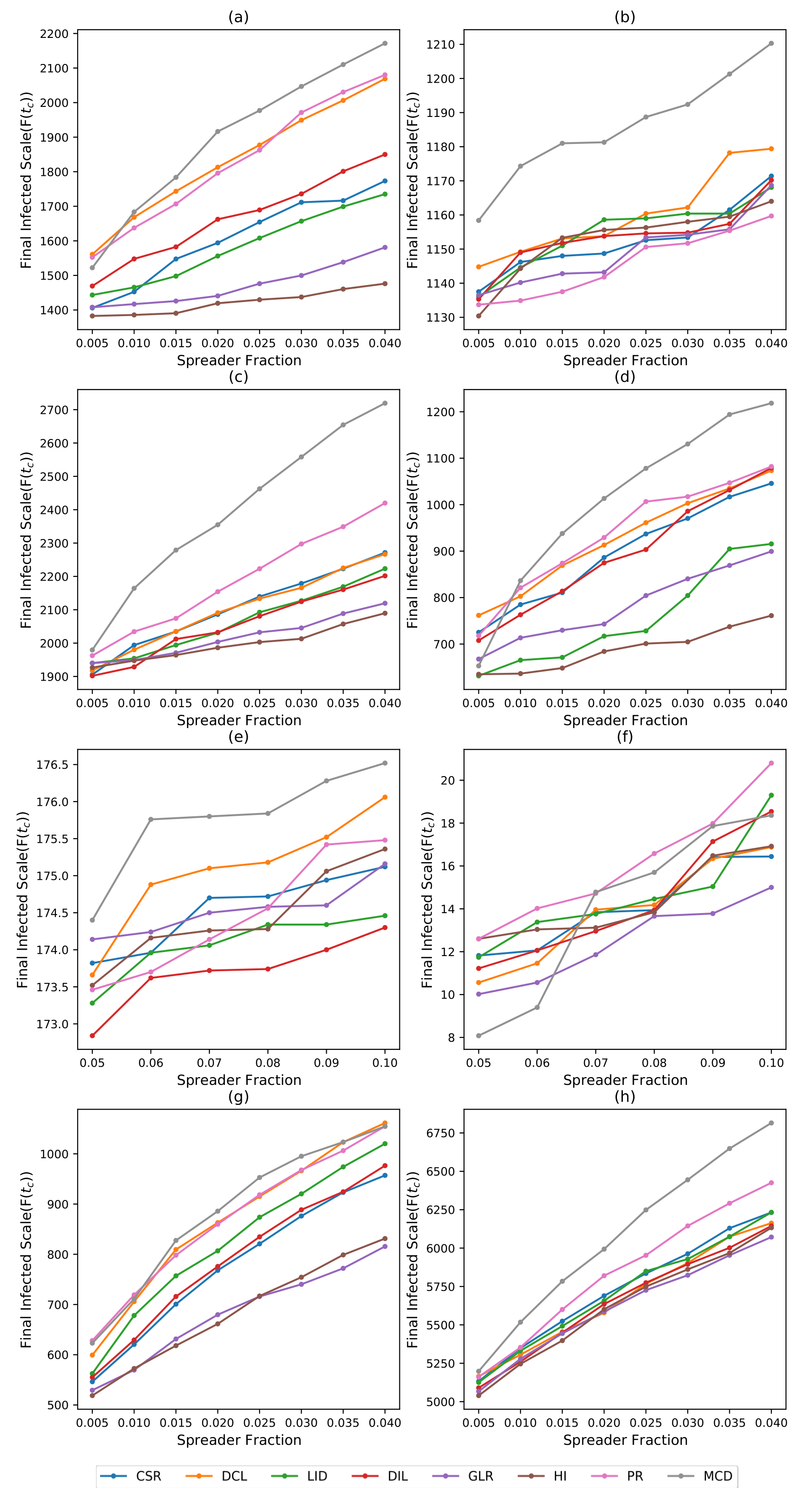}
	% figure caption is below the figure
	\caption{Final infected scale plots for various values for inital spreader fraction on the (a) PGP dataset (b) BA dataset (c) Ca-hepth dataset (d) GR-QC dataset (e) Jazz Musicians dataset (f) Dolphins dataset (g) Erdos992 dataset (h) p2p-Gnutella25 dataset. The results are averaged for 100 independent simulations of the Information Cascade model with an activation probability (P) equal to 0.1.}
	\label{FTC}       % Give a unique label      
\end{figure*}

The Final Infected Scale illustrates the final degree of propagation or \emph{infection} in the network at the end of the simulation, for varying values of fraction of nodes taken as initial spreaders. We evaluated our proposed approach against other competing approaches using the IC simulation model for 100 test runs and plotted the average results in Figure \ref{FTC}. For small datasets, i.e., datasets with the number of nodes less than 2000, we take the values of the initial spreader fraction from the set {0.02, 0.03, 0.04, 0.05, 0.06, 0.07, 0.08, 0.09, 0.1}, while for larger datasets with the number of nodes greater than 2000 nodes, we take the values of the initial spreader fraction from the set {0.005, 0.01, 0.015, 0.02, 0.025, 0.03, 0.035, 0.04}. We have taken larger values of the initial spreader fraction for smaller datasets, as the absolute number of initial spreaders will be small for smaller values of the initial spreader fraction in smaller datasets. We have taken the \emph{infection probability}, i.e., the probability of influencing a susceptible node as 0.1 for all combinations of datasets and approaches.

The graphs for Final Infected Scale with variations in Spreader Fraction for all approaches and datasets are illustrated in Figure \ref{FTC}. We observed that the final infected scale generally increased with an increase in spreader fraction, irrespective of the dataset. We observed that our proposed approach, MCD, outperformed competing approaches significantly across all datasets, except the Dolphins dataset, where PR outperforms our approach. We found DCL and PR to perform reasonably well across the majority of the datasets too. HI was consistently among the worst-performing approaches of the lot for most datasets. Thus, we conclude that our approach handily outperforms other approaches in maximizing information propagation in a network for a wide set of values of the initial spreader fraction.

\subsection{Average Distance between Spreaders}
It is desirable for the initial spreaders to have the least overlapping area of influence so that a wider portion of the network can be activated more efficiently. Thus, it is preferable that the average distance between spreaders is maximized. To compare our proposed MCD approach against other approaches, we calculated the average shortest distance between the initial spreaders ($L_s$) for a wide set of values of the spreader fraction.

The graphs for the Average Distance between Spreaders vs. Initial Spreader Fraction for all approaches and datasets are illustrated in Figure \ref{LS}. We observed that MCD significantly outperforms other approaches across datasets for almost all values of the initial spreader fraction. This ensures that MCD selects nodes that are far apart, thus avoiding overlapping of influence as stated in the manuscript. It verifies that MCD is able to cater to the rich club effect, which other competing approaches are unable to counter. The significant gap in the figure between our approach and competing approaches indicates that MCD selects nodes that are different from the ones selected by other approaches, and significantly farther apart. This further ensures nodes selected by MCD may not belong to the same region, thus having a wider spread of influence and lesser overlap between the regions of influence for each seed node. Further, we observe that on increasing the spreaders fraction, the difference between seed nodes decreases due to the increased absolute number of seed nodes for the whole network. This is observed in the case of all approaches, as increasing the absolute number of spreaders in a graph is bound to decrease the average distance between the spreaders. However, MCD still has a higher average spreader distance as compared to other approaches, even for greater values of Spreader Fraction, verifying MCD’s superiority over competing approaches. Among the other approaches, we found DCL and PR to perform reasonably well. However, there was still a significant gap between these approaches and MCD. We found GLR consistently performs the worst in this metric across almost all datasets. Thus, we conclude that MCD suffers from a significantly lesser \emph{rich club} effect and outperforms all other approaches significantly.

\subsection{Execution Time}

We compared our proposed approach, MCD, against other competing approaches based on the execution time required by each approach for ranking the nodes for all datasets. Figure \ref{ET} illustrates the line graph comparing MCD with other approaches based on the execution time for all datasets. DIL performed well on larger datasets but was relatively more computationally expensive for smaller datasets. We found LID and PR to be the least expensive computationally, while GLR was consistently the most expensive. Our proposed approach  was computationally the least expensive for smaller datasets, while it was more expensive than local hybrid centrality-based approaches like LID for larger datasets. Thus, we observe that MCD performs well in terms of execution time when compared to other approaches, especially when considering the improved performance in other metrics.

\section{Statistical Testing}
To further evaluate the performance of MCD using statistical tests, we have conducted the Friedman test \citep{Friedman}, which is a non-parametric statistical test used for multiple comparisons (more than two given methods). We have performed this test to determine whether the performance of our proposed method is significantly different than the other approaches for two or more data sets. The Friedman test attempts to detect significant differences based on the ranking of the methods, rather than their errors. It consists of two hypotheses: the null ($H_0$) and the alternate hypothesis ($H_1$). The former states that there are no prominent differences between these algorithms (equality of medians) and the latter insists that the algorithms have significant differences in their median of the population, thus negating the null hypothesis. We have performed the following test on the Final Infected Scale metric as well as the Average Distance between Spreaders metric.

In order to perform the Friedman Statistical Test, we execute the following procedure:

\begin{itemize}
    \item Gather the generated results for each problem pair.
    \item Rank the values in ascending order from 1 (best value) to $n$ (worst value) for each problem $i$ for a particular algorithm $j$.
    \item For algorithm $j$, the average rank for each problem $i$ is calculated using the Eq. \ref{RJ}:
    \begin{equation}\label{RJ}
        R_j = \frac{1}{n} \sum_{i=1}^n(r_i^j)
    \end{equation}
    where $r^j$ is the rank ($1<j<k$) and $R_j$ is the average rank.
    \item Now that all algorithms are ranked according to their priority, compute the Friedman Statistic $F_f$ using the following equation. $F_f$ is based on a Chi-square distribution, with $k-1$ degree of freedom. $F_f$ is computed as shown in Eq. \ref{FF}.
    \begin{equation}\label{FF}
        F_f = \frac{12n}{k(k+1)} \left[\sum_{j=1}^k R_j^2 - \frac{k(k+1)^2}{4}\right]
    \end{equation}
    \item The previous statistic produced relatively conservative results, which were not desired. Thus, \citep{Iman-Davenport} introduced a new statistic $F_{id}$, as shown below, which follows the $F-distribution$ with the degree of freedom as $k-1$ and $(n-1)(k-1)$, given in Eq. \ref{Eq:FID}:
    \begin{equation}\label{Eq:FID}
        F_{id} = \frac{(n-1)\chi_F^2}{n(k-1)-\chi_F^2}
    \end{equation}
\end{itemize}

Table \ref{FIS-statF} and Table \ref{LS-statF} show the average ranking calculated from the Friedman test for all the approaches used in this paper. MCD clearly has a better average ranking than the other competing approaches.

\begin{table}[!ht]
    \centering
    \caption{Average ranking of algorithms using the Friedman test based on Final Infected State}
    \begin{tabular}{|l|l|l|}
    \hline
        ~ & Method & Average Rank \\ \hline
        1 & MCD & 1.567  \\ \hline
        2 & PR & 2.983  \\ \hline
        3 & DCL & 3.467  \\ \hline
        4 & CSR & 4.675  \\ \hline
        5 & LID & 5.033  \\ \hline
        6 & DIL & 5.225  \\ \hline
        7 & GLR & 6.517  \\ \hline
        8 & HI & 6.533 \\ \hline
    \end{tabular}
    \label{FIS-statF}
\end{table}

\begin{table}[!ht]
    \centering
    \caption{Average ranking of algorithms using the Friedman test based on Average Distance between Spreaders}
    \begin{tabular}{|l|l|l|}
    \hline
        ~ & Approach  & Average Rank \\ \hline
        1 & MCD & 1.067  \\ \hline
        2 & PR & 2.983  \\ \hline
        3 & DCL & 3.517  \\ \hline
        4 & LID & 4.675  \\ \hline
        5 & HI & 5.3  \\ \hline
        6 & CSR & 5.367  \\ \hline
        7 & DIL & 5.508  \\ \hline
        8 & GLR & 7.583 \\ \hline
    \end{tabular}
\label{LS-statF}
\end{table}

The unadjusted P-value or Holm P-value, obtained from the Iman-Davenport statistic, corresponding to the performance of MCD, advocates the rejection of the null hypothesis $H_0$. All of the computed P-values are less than the standard significance level $\alpha = 0.05$. This indicates that there is a significant difference between the performance of the baseline approaches and our proposed approach. Hence, the obtained P-values conclude the negation of null hypothesis $H_0$ but are not appropriate for comparison with different methods. In order to compare these approaches with each other, we must calculate their respective adjusted P-values keeping MCD as the control algorithm.

Hence, we use adjusted P-values (APVs) for evaluation in order to perform a comparative study between the control method and benchmark methods on a statistical ground. Adjusted P-value provides the correct correlation between these algorithms, taking into account the accumulated family error with respect to the MCD control algorithm. Moreover, the adjusted P-values can be directed compared with the significance level $\alpha$, which is equivalent to 0.05. In order to calculate the adjusted P-values, a few post-hoc procedures need to be defined. Various post-hoc procedures such as \citep{bonn} and \citep{holland} differ as they adjust the value of $\alpha$ to compensate for multiple comparisons for multiple methods. In this paper, we have used the common Holm's procedure \citep{Holm} to evaluate the respective APVs. The values of Holm P-values and APVs are always sorted in ascending order. The equation for the same is given below, where indices $i$ and $j$ refers to the main hypothesis whose APVs are being computed and different hypothesis in the set respectively. $P_j$ is the P-value for the $j^{th}$ hypothesis. Holm APV is computed as shown in Eq. \ref{Eq:Holm}.

\begin{equation}\label{Eq:Holm}
    Holm APV_i = \min\{v,1\}, \text{where} \,v = \max\{(k-j)p_j:1\leq j\leq i\}
\end{equation}

\begin{table}[!ht]
    \centering
    \caption{Adjusted P-values (APVs) using Holm procedure for Final Infected State}
    \begin{tabular}{|l|l|l|l|l|}
    \hline
        ~ & Approach & Z-score & P-value & APV \\ \hline
        1 & HI & -11.11 & 5.88E-29 & 4.11E-28  \\ \hline
        2 & GLR & -11.06 & 8.91E-29 & 5.35E-28  \\ \hline
        3 & DIL & -8.18 & 1.42E-16 & 7.08E-16  \\ \hline
        4 & LID & -7.75 & 4.53E-15 & 1.81E-14  \\ \hline
        5 & CSR & -6.95 & 1.82E-12 & 5.46E-12  \\ \hline
        6 & DCL & -4.24 & 1.08E-05 & 2.15E-05  \\ \hline
        7 & PR & -3.16 & 7.68E-04 & 7.68E-04 \\ \hline
    \end{tabular}
    \label{FIS-statH}
\end{table}

\begin{table}[!ht]
    \centering
    \caption{Adjusted P-values (APVs) using Holm procedure for Average Distance between Spreaders}
    \begin{tabular}{|l|l|l|l|l|}
    \hline
        ~ & Approach & Z-score & P-value & APV \\ \hline
        1 & GLR & -14.57 & 2.13E-48 & 1.49E-47  \\ \hline
        2 & DIL & -9.93 & 1.51E-23 & 9.08E-23  \\ \hline
        3 & CSR & -9.61 & 3.45E-22 & 1.73E-21  \\ \hline
        4 & HI & -9.46 & 1.45E-21 & 5.81E-21  \\ \hline
        5 & LID & -8.06 & 3.56E-16 & 1.07E-15  \\ \hline
        6 & DCL & -5.47 & 2.15E-08 & 4.29E-08  \\ \hline
        7 & PR & -4.28 & 9.10E-06 & 9.10E-06 \\ \hline
    \end{tabular}
\label{LS-statH}
\end{table}

The results of the APVs using Holm's procedure are shown in Table \ref{FIS-statH} and Table \ref{LS-statH}. The intended APVs are less than the significance level, thus rejecting the null hypothesis. It is clearly observed from the results obtained that MCD is significantly better than the competing approaches on these statistical tests as well.

\section{Conclusion}
Finding influential nodes in complex networks like social networks play a vital role in achieving maximum information spread across the network. It has found numerous applications in business, marketing, and generating awareness for a social cause. In this work, we proposed a modified Community Diversity (MCD) method for finding influential nodes. First, we introduced the concept of Extended Community Diversity, which represents the community diversity score of the extended neighborhood of a node. Next, we introduced Modified Community Diversity, built upon the concepts of Community Diversity and Extended Community Diversity. To verify the efficacy of our approach in determining the most effective influential nodes in a network, we compared MCD with seven recent state-of-the-art methods over eight datasets. We found that MCD cumulatively outperformed the competing approaches across three metrics. Further, we observed that seed nodes selected by MCD were reasonably close together for extended activation due to a node’s influence but far enough to not significantly overlap with each other’s scope of influence, thus mitigating the Rich Club effect significantly.

\bibliographystyle{cas-model2-names}

\bibliography{main}

\end{document}